\documentclass[iop]{emulateapj}

\newif\ifpreprint
\preprinttrue
\preprintfalse

\usepackage{xspace}
\usepackage{color}
\usepackage{enumitem}
\usepackage{hyperref}

\newcommand{\lya}        {Ly$\alpha$\xspace}

\newcommand{\unitcgssb}  {erg\,s$^{-1}$\,cm$^{-2}$\,arcsec$^{-2}$\xspace}

\newcommand{\unitcgslum} {erg\,s$^{-1}$\xspace}
\newcommand{\msun}       {M$_{\odot}$}
\newcommand{\ergs}       {erg s$^{-1}$}
\newcommand{\Llya}       {$L_{\rm Ly\alpha}$\xspace}
\newcommand{\pol}        {\ifmmode{P_{\%}}\else$P_{\%}$\xspace\fi}
\newcommand{\pollya}     {$P_{\%, {\rm Ly}\alpha}$\xspace}
\newcommand{\polcont}    {$P_{\%, {\rm cont}}$\xspace}

\newcommand{\kms}        {\ifmmode{\rm \,km\,s^{-1}}\else\,km\,s$^{-1}$\xspace\fi}
\newcommand{\um}         {$\mu$m\xspace}

\newcommand\ra[3]        {#1$^{\rm h}$\,#2$^{\rm m}$\,#3$^{\rm s}$}
\newcommand\dec[3]       {#1\degr\,#2\arcmin\,#3\arcsec}

\slugcomment{Accepted for publication in ApJ.}
\shorttitle{Polarization of the Radio-Loud \lya Nebula}
\shortauthors{You et al.}

\begin{document}

\title{Mapping the Polarization of the Radio-Loud \lya Nebula B3 J2330+3927\altaffilmark{*}}

\author{Chang You\altaffilmark{1},
        Ann Zabludoff\altaffilmark{1},
        Paul Smith\altaffilmark{1},
        Yujin Yang\altaffilmark{2,3,4,$\dagger$},
        Eunchong Kim\altaffilmark{5}, \\
        Buell Jannuzi\altaffilmark{1},
        Moire K. M. Prescott\altaffilmark{6,7},
        Yuichi Matsuda\altaffilmark{8,9},
        Myung Gyoon Lee\altaffilmark{5}\medskip}

\affil{$^{1\,}$Steward Observatory, University of Arizona, 933 N Cherry Ave., Tucson, AZ 85721}
\affil{$^{2\,}$Korea Astronomy and Space Science Institute, 776 Daedeokdae-ro, Yuseong-gu, Daejeon 34055, Korea}
\affil{$^{3\,}$Argelander Institut f\"ur Astronomie, Universit\"at Bonn, Auf dem H\"ugel 71, 53121 Bonn, Germany}
\affil{$^{4\,}$Korea University of Science and Technology (UST), 217 Gajeong-ro Yuseong-gu, Daejeon 34113, Korea}
\affil{$^{5\,}$Department of Physics and Astronomy, Seoul National University, Gwanak-gu, Seoul 88226, Korea}
\affil{$^{6\,}$Department of Astronomy, New Mexico State University, 1320 Frenger Mall, Las Cruces, NM 88003, USA}
\affil{$^{7\,}$Dark Cosmology Centre, Niels Bohr Institute, University of Copenhagen, Juliane Maries Vej 30, DK-2100 Copenhagen \O, Denmark}
\affil{$^{8\,}$National Astronomical Observatory of Japan, National Institutes of Natural Sciences, 2-21-1 Osawa, Mitaka, Tokyo 181-8588, Japan}
\affil{$^{9\,}$The Graduate University for Advanced Studies (SOKENDAI), 2-21-1 Osawa, Mitaka, Tokyo 181-8588, Japan\medskip}

\altaffiltext{*}{Observations reported here were obtained at the MMT
Observatory, a joint facility of the University of Arizona and the
Smithsonian Institution.}

\altaffiltext{$\dagger$}{To whom correspondence should be addressed.
Email address: yyang@kasi.re.kr}


\begin{abstract}
\lya nebulae, or ``\lya blobs'',  are extended (up to $\sim$100\,kpc),
bright (\Llya\ $\gtrsim 10^{43}$ \ergs) clouds of \lya emitting gas
that tend to lie in overdense regions at $z$ $\sim$ 2--5.  The origin
of the \lya emission remains unknown, but recent theoretical work
suggests that measuring the polarization might discriminate among
powering mechanisms.  Here we present the first narrowband, imaging
polarimetry of a radio-loud \lya nebula, B3 J2330+3927 at $z=3.09$,
with an embedded active galactic nucleus (AGN).  The AGN lies near
the blob's \lya emission peak and its radio lobes align roughly
with the blob's major axis.  With the SPOL polarimeter on the 6.5m
MMT telescope, we map the total (\lya + continuum) polarization in a
grid of circular apertures of radius 0.6$^{\prime\prime}$ (4.4\,kpc),
detecting a significant ($>2\sigma$) polarization fraction \pol in nine
apertures and achieving strong upper-limits (as low as 2\%) elsewhere.
\pol increases from $< 2$\% at $\sim$5\,kpc from the blob center to 17\%
at $\sim$15--25\,kpc.  The detections are distributed asymmetrically,
roughly along the nebula's major axis.  The polarization angles $\theta$
are mostly perpendicular to this axis.  Comparing the \lya flux to that
of the continuum, and conservatively assuming that the continuum is
highly polarized (20--100\%) and aligned with the total polarization,
we place lower limits on the polarization of the \lya emission \pollya
ranging from no significant polarization at $\sim$5 kpc from the blob
center to 3--17\% at 10--25\,kpc.  Like the total polarization, the \lya
polarization detections occur more often along the blob's major axis.
\end{abstract}


\section{Introduction}

Giant (up to $\sim$100\,kpc) gaseous, \lya-emitting nebulae, also known
as \lya ``blobs" \citep{Steidel2000, Matsuda2004, Dey2005, Prescott2008,
Yang2009}, are extremely luminous (\Llya\ $\gtrsim 10^{43}$ \ergs)
and were discovered first in overdense regions of the high redshift
($z$ $\sim$ 2--5) Universe \citep{Matsuda2005, Prescott2008}. Their
rarity and clustering are consistent with their occupying massive
($\sim$10$^{13}$\,\msun) dark matter halos then that evolve into rich
groups or clusters of galaxies today \citep{Yang2009,  Yang2010}.
The \lya blob gas thus may represent the proto-intracluster medium and
the embedded sources the progenitors of cluster galaxies \citep{Yang2010,
Prescott2012}.  Thus identifying the mysterious source or sources of
the extended \lya emission is essential to understanding the evolution
of large-scale structure and of the most massive galaxies.

Observations and theory suggest a range of powering mechanisms,
including gravitational cooling radiation \citep{Haiman2000, Fardal2001,
Goerdt2010, Faucher-Giguere2010,  Rosdahl-Blaizot2012}, shock-heating
from starburst-driven winds \citep{Taniguchi-Shioya2000, Mori2004},
the resonant scattering of \lya photons produced by star formation
\citep{Steidel2011}, and photo-ionizing radiation from active galactic
nuclei (AGN) \citep{Haiman2000}.  Even with careful constraints on the
\lya line profile and distribution, discriminating among these models
is difficult, in part due to the complex radiative transfer of the
resonantly scattered \lya line and the uncertain internal geometry of
each \lya blob \cite[e.g.,][]{Yang2011,Yang2014a,Yang2014b}.

Measuring the polarization of the \lya line can shed new light on the
problem.  Recent radiative transfer simulations predict the polarization
of the \lya line in a number of different scenarios.  For example,
backscattered \lya flux from galaxies surrounded by a superwind-driven
outflow is expected to produce a \lya polarization fraction \pollya
that rises with radius to as much as $\sim$40\% where the neutral
hydrogen column density $N_{\rm HI}$ drops below $10^{19}$ cm$^{-2}$
\citep{Dijkstra-Loeb2008}.  A similar \pollya integrated over the line
profile may arise from cooling radiation from a collapsing proto-galaxy
(\citealt{Dijkstra-Loeb2008}; see also \citealt{Trebitsch2016}), but with
the inverted wavelength dependence when the line is spectrally resolved.
Resonant scattering in the diffuse intergalactic medium typically results
in a lower \pollya ($\sim$7\%), which depends on the flux of the ionizing
background \citep{Loeb&Rybicki1999, Rybicki&Loeb1999, Dijkstra-Loeb2008}.
These models, which all currently assume spherical symmetry, continue to
grow more sophisticated \cite[e.g.,][]{Trebitsch2016}.  Their improving,
detailed predictions, when combined with the new availability of
polarimeters on the largest telescopes, provide a unique opportunity to
isolate the mechanism that powers \lya blobs by mapping the polarization.

Polarization work on \lya blobs has been limited.  To date, only two
\lya blobs have been observed with narrowband imaging polarimetry.  One,
SSA22--LAB1, shows concentric polarization rings, reaching $\sim$10\%
at $\sim$30\,kpc from the blob center and rising to $\sim$20\% at
$\sim$45\,kpc \citep{Hayes2011}, suggesting a central powering source for
this \lya blob \cite[see also][]{Beck2016}.  In the other blob, LABd05,
\citet{Prescott2011} do not detect polarization within a single, large
(radius $\sim$ 33 kpc) aperture, obtaining an upper-limit of 2.6\% $\pm$
2.8\%; deeper and spatially resolved observations are required to test
this result (E.~Kim et al, in preparation).  These past studies assume
that the polarization arises solely from \lya, given that the \lya line
dominates the continuum emission, at least at large radii.  Both \lya
nebulae are radio-quiet.  Spectro-polarimetry of a radio-loud \lya nebula,
TXS 0211--122 at $z=2.3$, reveals polarization of the \lya line: 16.4\%
$\pm$ 4.6\% on one side of the nebula \citep{Humphrey2013}.  In this
case, the spatial information is limited, inhibiting the interpretation
of the results.

Looking to the literature on radio galaxies, which can be surrounded by
line emission nebulae similar in \lya luminosity and spatial extent to
blobs \cite[see][and references therein]{McCarthy1993} does not improve
our understanding of how the \lya polarization is distributed on the sky.
Existing polarization measurements of radio galaxies, seeking to explain
the alignment effect---the strong correlation between their radio
and optical continuum morphologies \citep{McCarthy1987}---tend to focus 
on the continuum \citep{Vernet2001}.
Constraints on the \lya polarization are few, particularly over the
tens of kpc scales typical of \lya blobs.  Using spectro-polarimetry,
\citet{Cimatti1998} find that the \lya line is unpolarized in two radio
galaxies.  \lya around another radio galaxy, 4C 41.1, is polarized
at a low level (1.12\% $\pm$ 0.26\%), while its continuum emission is
unpolarized \citep{Dey1997}.


The similarity in morphology and energy between extended \lya nebulae with
radio-loud and radio-quiet AGN suggest an unexplored connection between
their powering mechanisms \citep{Villar-Martin2003, Dey1997}.  Here we
present the first \lya imaging polarimetry measurement for a blob with
an embedded radio galaxy.  We use the SPOL imaging spectro-polarimeter
on the 6.5m MMT telescope to map B3 J2330+3927, a radio-loud \lya blob
at $z=3.087$.  Its embedded radio galaxy is one of the 1103 radio sources
from the Third Bologna Catalog \citep{Ficarra1985, Vigotti1989}.
The associated \lya nebula was discovered by \citet{DeBreuck2003} through
a long-slit spectroscopy and observed in detail by \cite{Matsuda2009}.
SPOL is a clean instrument, designed to reduce any instrument polarization
by integrating over 16 different waveplate positions.  At the redshift
of our source, SPOL's high stability and sensitivity on the MMT enables
measurement of a few percent polarization on scales of $\sim$5\,kpc,
even at the low surface brightnesses characteristic of \lya blobs.

This paper is the first of several to map the polarization of giant
\lya nebulae at high-redshift. In this paper, we present the map of our
first target and establish our methodologies.  Subsequent papers will
analyze the full blob sample and compare the results to physical models.
This paper is organized as follows.  In Section \ref{sec:obs}, we describe
the details of our observations. In Section \ref{sec:data}, we discuss
the data reduction for the polarization measurement and the calibration
sources. In Section \ref{sec:results}, we present our polarization map and
discuss the possible sources of error. In Section \ref{sec:conclusion},
we summarize our conclusions.

\section{The Observations}
\label{sec:obs}

\subsection{The Target}

B3 J2330+3927 is a high-redshift ($z=3.087$) radio galaxy at
R.A.=\ra{23}{30}{24.9} and decl.~=\dec{39}{27}{12} that is embedded in
a giant Ly$\alpha$ halo that extends over $\sim$130\,kpc.  This nebula
is one of the brightest known, with \Llya = 2.5$\times$ $10^{44}$
\unitcgslum \citep{Matsuda2009}. The CO emission and \ion{H}{1}
absorption reveals a massive gas and dust reservoir associated with
the radio galaxy \citep{DeBreuck2003, Ivison2012}.  VLBA and VLA data
show a one-sided jet driven by a Type II AGN \citep{Perez-Torres2005}.
The galaxy environment of this \lya blob is over-dense:  a combination of
broad and narrowband observations \citep{Matsuda2009} reveals 127 compact
Ly$\alpha$ emitter (LAE) candidates and another giant ($\sim$100\,kpc),
but radio-quiet, \lya blob within the $31^{\prime} \times 24^{\prime}$
(58 $\times$ 44 comoving Mpc$^2$) field. This wealth of ancillary data,
the redshift, and a bright point source at R.A.=\ra{23}{30}{25.10},
decl.~=\dec{39}{27}{05.4} useful for image registering and alignment,
make B3 J2330+3927 an attractive target.

\subsection{The Instrument}
\label{sec:instrument}

On UT September 18--20, 2012, we used the 6.5m MMT telescope
on Mount Hopkins, Arizona, to observe B3 J2330+3927 with the
SPOL CCD imaging/spectro-polarimeter in its imaging polarimetry mode
\citep{Schmidt1992b}.  We used a narrowband filter ({\tt kp583}) on loan
from Kitt Peak National Observatory that is centered at 4980\,\AA\ and
has a FWHM of 54\,\AA.  The detector is a thinned, anti-reflection-coated
1200$\times$800 STA CCD with a pixel scale of 0\farcs19 per pixel and a
quantum efficiency of $\sim$0.85 in the filter bandpass.  We obtained a
total of 8.6 hours exposure time on B3 J2330+3927.  For the calibration
of the instrument, we observed unpolarized and polarized standard stars
each night.  We also observed CRL 2688 (the ``Egg Nebula'') as an extended
polarized source to investigate any unforeseen systematic effects across
the 19\arcsec$\times$19\arcsec\ field of view.


In SPOL, the telescope is fed through a half-wave plate and then to
a Wollaston prism.  The Wollaston prism is located in the optical path
between a transmissive collimator and a plane mirror that substitutes for
a diffraction grating when imaging polarimetry is desired.  The narrowband
filter is placed in the collimated beam between the collimator and the
Wollaston prism.  The half-wave plate retards one orthogonal component of
the light and thus changes the polarization angle of the incoming light.
The Wollaston prism splits the two orthogonal polarizations so the two
polarizations are imaged separately, in our case in separate ``panels''
in one ``image''. The difference between these two panels indicates the
strength of the polarization.

Linear polarization measurements with SPOL are accomplished by stepping
a wheel holding a semi-achromatic half-wave plate through two sequences
that are aimed to measure Stokes parameters $Q$ and $U$, respectively.
A $Q$-sequence yields two images ($Q^+$ and $Q^-$): the first ($Q^+$)
consisting of two beams (panels) of four exposures at four orthogonal
position angles of the waveplate (0\degr, 90\degr, 180\degr, 270\degr).
The second image is taken at angles offset by 45 degrees from the first
(45\degr, 135\degr, 225\degr, 315\degr).
The $U$-sequence follows the same progression ($U^+$ and $U^-$) as
the $Q$-sequence, but the waveplate position angles are offset by 22.5
deg from those of the $Q$-sequence.  The redundancy in the data-taking
sequences ensures that effects caused by imperfections in the waveplate
and the waveplate's positioning in the optical path are minimized.  As a
result, the instrumental polarization of SPOL is consistently $<0.1\%$
and verified by our measurements of unpolarized standards during the
nights that we observed B3 J2330+3927 (Section \ref{sec:standard}).
We do not include this negligible polarization in the subsequent analysis
of the data.  In addition, the dual-beam design of SPOL eliminates the
possibility of measuring spurious polarization arising from variable
seeing and sky transparency during observing sequences.

For B3 J2330+3927, we took exposures of 300 sec per waveplate position
angle, so we completed both $Q$ and $U$ sequences in 80 min.  In total,
we obtained six full polarization sequences.  We optimized the MMT optics
between each measurement, except when the seeing remained ideal and the
weather conditions did not change. The seeing was $\sim$1.0\arcsec\
during most of the observations, rising to 1.5\arcsec\ for the two
sequences taken at the end of each of the two nights.
We used the positions of the \lya blob center (R.A.=\ra{23}{30}{24.9},
decl.~=\dec{39}{27}{12}) and of a bright point source
(R.A.=\ra{23}{30}{25.10}, decl.~=\dec{39}{27}{05.4}; $\sim$8\arcsec\
to the southeast of the blob center) to register and align our images,
as the field was dithered slightly between polarization sequences to
minimize the effects of any poorly calibrated pixels.
We measured the polarization efficiency of the system ($p_{\rm eff}$
$\approx$ 0.973) by inserting a Nicol prism before the aperture plate and
waveplate in the light path within the instrument.  This efficiency is
consistent with other measurements acquired over more than two decades
for SPOL at 4980\AA\ when used as a spectro-polarimeter.

\begin{figure}
\epsscale{1.2}
\ifpreprint\epsscale{0.90}\fi
\plotone{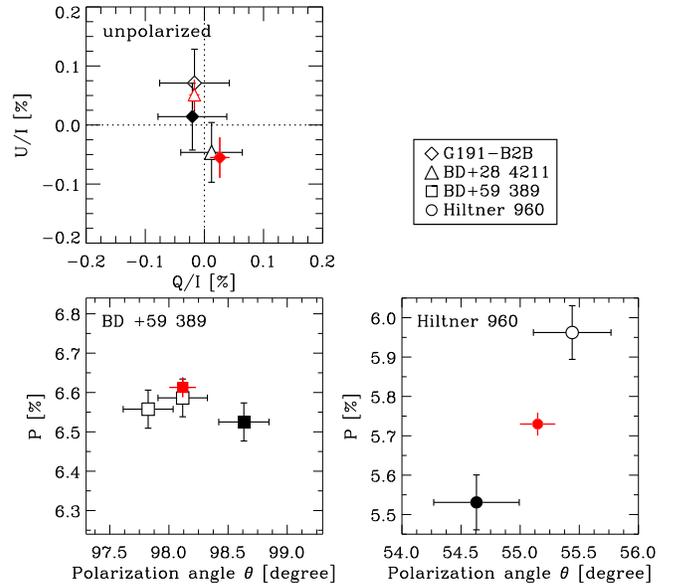}
\caption{
Comparison of our standard star polarization measurements (black symbols)
with values derived from the literature (red symbols). Black open and
filled symbols represent the first and second night of observations.
Different symbol shapes indicate the different stars. The error bars
represent our $1\sigma$ photon noise.
{\bf Top left:}  Our $Q/I$ and $U/I$ measurements for unpolarized
standard stars. All three measurements are consistent with zero and the
measurements in the literature.
{\bf Bottom left:} \pol and $\theta$ measurements for the polarized star
BD+59 389.  All three measurements (black) are within $\sim$2$\sigma$
of the expected \pol and $\theta$ (red).
{\bf Bottom right:} Results for the polarized star Hiltner 960.
The measured \pol's are more discrepant ($\sim$3.1$\sigma$) than for BD+59
398, likely due to the variability of Hiltner 960.  Note that this
error ($\Delta$\pol\ $\lesssim$ 0.2\%) is negligible compared to the
uncertainties in the measurements of the science target, B3 J2330+3927.
}
\label{fig:polcal}
\end{figure}

\section{The Data Reduction}
\label{sec:data}

\subsection{Pre-processing}

To prepare the images for polarization measurement, we perform overscan
correction, bias subtraction, flat fielding, and cosmic-ray removal.

For flat-fielding, we obtained dome flats with all the polarization
optics (Wollaston prism and half-wave plate) and construct a partial
skyflat by median-combining the science images with the central \lya
blob blocked out.  The dome flats show a significant gradient across the
image that the science exposures and partial skyflat do not.  To correct
the domeflat, we fit the gradient with a 2-D first-order polynomial
and divide it out.  We apply the resulting ``flattened" dome flat to
the partial skyflat and the science images.  There are no significant
gradients in the resulting images.

We use the {\tt L.A.COSMIC} package \citep{vanDokkum2001} to remove
cosmic rays from our images. We examine the cosmic ray masks by eye to
confirm that real signal from the nebula remains.

\subsection{Polarization Calculation}

As described in \ref{sec:instrument}, from a full $Q$--$U$ sequence,
we obtain a total of four images ($Q^+$, $Q^-$, $U^+$, $U^-$), each with
two panels (``up'' and ``down'' beams), respectively.  Here we explain
the calculation of the polarization parameters from those images.

With the notation
\begin{equation}
	q\equiv\frac{Q}{I} {\rm ~~and~~} u\equiv \frac{U}{I},
\end{equation}
$q$ and $u$ can be determined using the following formulae:
\begin{equation}
\label{eq:q1}
q =  \frac{Q}{I_Q} = 
 \frac{1}{2}\left[\left(\frac{Q^{-} - Q^{+}}{Q^{-} + Q^{+}}\right)_{\rm up}\! 
                + \left(\frac{Q^{+} - Q^{-}}{Q^{+} + Q^{-}}\right)_{\rm down}\right]
\end{equation}
\begin{equation}
\label{eq:q2}
u =  \frac{U}{I_U} = 
 \frac{1}{2}\left[\left(\frac{U^{-} - U^{+}}{U^{-} + U^{+}}\right)_{\rm up}\! 
                + \left(\frac{U^{+} - U^{-}}{U^{+} + U^{-}}\right)_{\rm down}\right],
\end{equation}
where ${I_Q}$ and ${I_U}$ are the total intensities measured
from the $Q$ and $U$ sequences, respectively:
\begin{eqnarray}
I_Q  &~=~&  [ {(Q^{-} + Q^{+})}_{\rm up} + {(Q^{-} + Q^{+})}_{\rm down} ]/2 \\
I_U  &~=~&  [ {(U^{-} + U^{+})}_{\rm up} + {(U^{-} + U^{+})}_{\rm down} ]/2 \\
I    &~=~&  \frac{1}{2}(I_Q + I_U).
\end{eqnarray}
Ideally, $Q^+_{\rm up}$ is the same as $Q^-_{\rm down}$ and $Q^+_{\rm
down}$ is the same as $Q^-_{\rm up}$.  The same applies for the $U$
images.

For each $Q$--$U$ sequence, we create these $I_i$, $Q_i$ and $U_i$ images
(or $q_i$ and $u_i$), and combine them to increase the signal-to-noise
(S/N).  When combining the sequences, we scale the images to compensate
for the variations arising from airmass and weather.  From these
final Stokes images ($I$, $Q$, $U$), we calculate the polarization
fraction (\pol) and angle ($\theta$) using the following formulae:
\begin{eqnarray}
   {\pol} &=& \sqrt{q^2 + u^2} \\
   \theta &=& \frac{1}{2} \arctan{\frac{U}{Q}}.
\end{eqnarray}
Because the S/N of our target is low, we calculate \pol and $\theta$
for large aperture sizes (1.2--1.5\arcsec) over the map. The error
associated with the polarization due to photon noise is derived from
propagating errors through the above formulae.

\subsection{Calibrations}

\subsubsection{Standard Stars}
\label{sec:standard}

To calibrate and verify the  linear polarization measurements with
SPOL, we observed both polarized and unpolarized standard stars
\citep{Schmidt1992a} each night.  These observations are summarized
in Figure \ref{fig:polcal}.  For the unpolarized stars, G191-B2B and
BD+28 4211, the instrumental polarization ($Q/I$ and $U/I$) at the MMT
is indeed $<0.1\%$ (top left panel), as previously found for SPOL at
other telescopes.  We also use these spectro-photometric standard stars
to flux-calibrate the narrowband images.

We observed two interstellar polarized standards, BD+59 389 and Hiltner
960.  Given that our narrowband filter is centered at a different
wavelength (4980 \AA) than previous measurements of these standards,
we calculate the expected \pol and $\theta$ within our bandpass by
interpolating between the previous measurements with an interstellar
polarization function \citep{Serkowski1973}.  Our observations of
BD+59 389 are consistent with historical measurements, i.e., our
three measurements over two nights agree within $\pm$1.6$\sigma$
and $\pm$2.0$\sigma$ of the interpolated \pol and $\theta$ from the
literature, respectively,

For Hiltner 960, the observed $\theta$'s are also within
$\pm$\,1.3$\sigma$ range, but the \pol's are more discrepant
($\sim$3.1$\sigma$) from the value derived from the literature.  Possible
reasons for this discrepancy include Hiltner 960's variability, leading
to a poorly-fit interstellar polarization curve \citep{Schmidt1992a},
and its close companion, which cannot be easily accounted for in the
polarization measurement.  Regardless of its source, this discrepancy
($\Delta$\pol\ $\lesssim$ 0.2\%) is negligible compared with the errors
in \pol and $\theta$ arising from photon noise in the measurement of
our science target, B3 J2330+3927.

\subsubsection{Egg Nebula}

\begin{figure}
\epsscale{0.95}
\ifpreprint\epsscale{0.75}\fi
\plotone{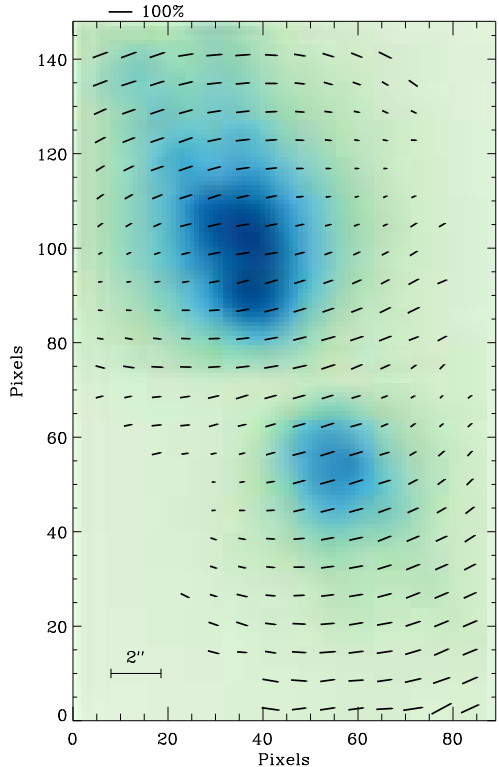}
\caption{ 
The SPOL polarization map of the extended polarization standard, CRL 2688
(the ``Egg Nebula'').  Our polarization vectors (all detected at $\geq
2\sigma$) for the northern and southern lobes are superposed on the total
flux map.  The FOV is 17\arcsec $\times$ 28\arcsec.  Each vector box is
1.14\arcsec $\times$ 1.14\arcsec, which matches the FWHM of our seeing.
The observed polarization patterns are qualitatively consistent with the
NIR polarization map from \citet{Sahai1998} in the sense that the vectors
along the axis connecting the two lobes are generally perpendicular to it.
}
\label{fig:eggmap}
\end{figure}

In addition to the polarization standard stars, we observe CRL 2688
as an extended polarization ``standard'' (Figure \ref{fig:eggmap}).
The short ($2\times960$ sec) observations of both the north and south
lobes test the polarization characteristics of SPOL over the entire field
of view.  We use these high S/N data to examine the images for unexpected
systematic effects that would be hidden in the case of a nebula as faint
as B3 J2330+3927.  Our optical polarization map is roughly consistent with
the NICMOS 2\um polarization map from \cite{Sahai1998}, e.g., the vectors
along the axis connecting the two components are generally perpendicular
to it (see their Fig.~5).  Furthermore, our average \pol for each lobe
lies within $1\%$ of the value expected at 4980\,\AA, as interpolated
from the optical polarization measurements of \citet{Shawl-Tarenghi1976}.

\section{Results and Discussion}
\label{sec:results}

\subsection{Total Polarization Map}
\label{sec:pol_total}

\begin{figure}
\epsscale{1.1}
\ifpreprint\epsscale{0.90}\fi
\centering
\plotone{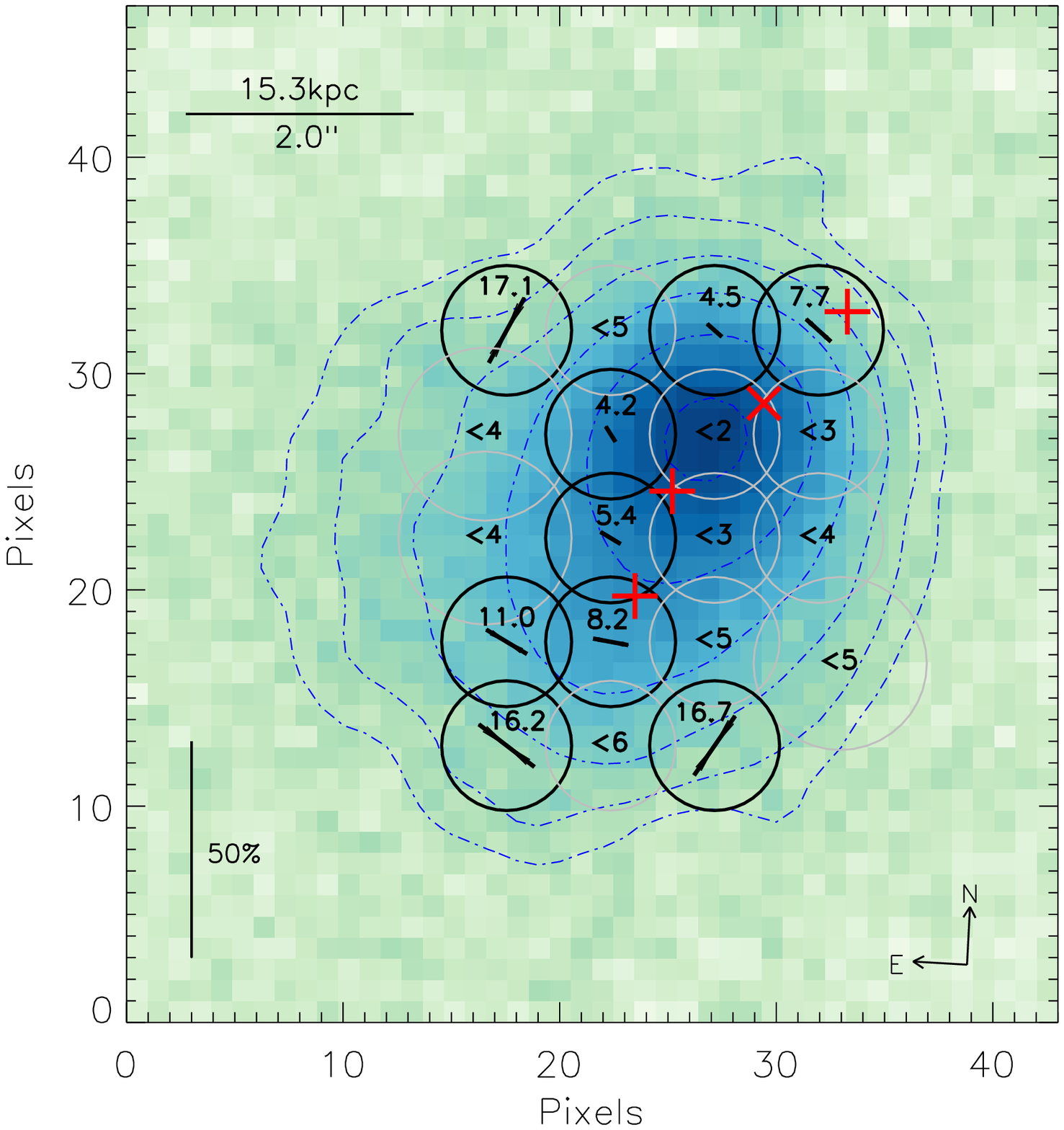}\\
\caption{
Total (\lya plus continuum) polarization map overlaid on the \lya flux
map for B3 J2330+3927.  The blue dot-dashed contours are 0.5, 1, 2,
4, 8, 16 and 32$\times10^{-17}$ \unitcgssb.  The red symbols show the
locations of the radio galaxy ($\times$) and the three radio knots ($+$)
in its lobes \citep{DeBreuck2003, Perez-Torres2005} with astrometric
uncertainties of $\sim$0.5\arcsec\ relative to the optical frame.
Any \pol measured at $\geq 2\sigma$ within an aperture is listed in that
aperture and shown as a vector.  We also plot the $\pm1\sigma$ errors in
\pol (vector length) and in $\theta$ (vector direction) derived from the
photon noise.  For apertures without significant ($\geq 2\sigma$) \pol
detections, we list the $2\sigma$ upper-limit.  The nine significant
polarization detections lie primarily along the blob's major axis
(also the radio lobe axis) and have angles perpendicular to it, with
\pol generally increasing further from blob center (and AGN).
}
\label{fig:polmap}
\end{figure}

Figure~\ref{fig:polmap} shows our polarization map of B3 J2330+3927 for
the light in the narrowband image,  i.e., \lya plus continuum centered
at 4980\,\AA\ with a FWHM of 54\,\AA.  We measure the polarization on a
grid of circular apertures with minimum radius of $R = 3$ pixels (i.e.,
$0.6^{\prime\prime}$, 4.4\,kpc), comparable to the seeing.  We enlarge
three apertures far from the \lya peak from $R = 3$ pixels to 4 pixels,
so that they reach a similar flux signal-to-noise ratio as the other
apertures.

\begin{figure}
\epsscale{1.25}
\ifpreprint\epsscale{1.00}\fi
\centering
\hspace{-0.7cm}
\plotone{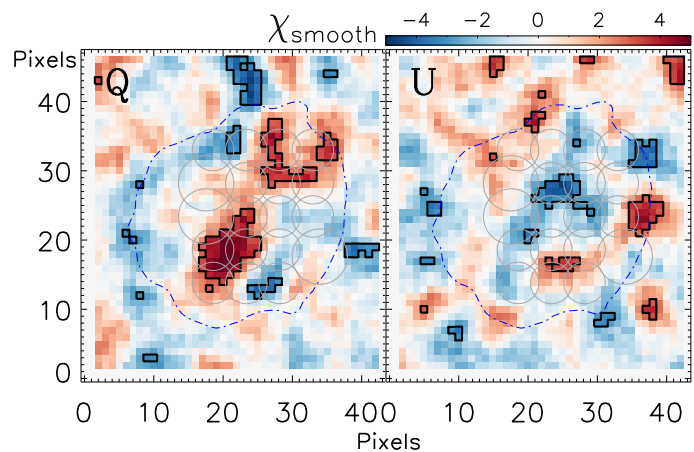}
\caption{
The $\chi_{\rm smooth}$ maps for the $Q$ and $U$ fluxes. The black
solid lines outline the pixels with $|\chi_{\rm smooth}| > 3$, i.e.,
statistically significant regions. The dot-dashed contour is the
outermost contour in Figure \ref{fig:polmap}.  The region with $|\chi_{\rm
smooth}| > 3$ in the $Q$ image is roughly aligned with the major axis,
demonstrating the significance of our polarization detections.
}
\label{fig:polmap_chi}
\end{figure}

We detect significant ($\geq$\,2$\sigma$) polarization in nine apertures
and achieve strong upper-limits (i.e., \pol as low as 2\%) elsewhere,
indicating varying polarization across the blob.  There is little if
any polarization at the blob center and to the southwest of the nebula.
The significant detections are generally distributed along the blob's
major axis, which is also the radio lobe direction.  Along that axis,
\pol increases from $< 2\%$ at $\sim$5\,kpc from the blob center to
roughly 17\% at $\sim$15--25\,kpc.  The polarization angles tend to be
perpendicular to that axis.

To test the significance of our polarization measurements, we
show the smoothed-$\chi$ images for $Q$ and $U$ fluxes in Figure
\ref{fig:polmap_chi}. Here $\chi_{\rm smooth}$ of an image $I$ is
defined by
\begin{equation}
\chi_{\rm smooth} = \frac{I_{\rm smooth}}{\sigma_{\rm smooth}} 
                  = \frac{I_i \ast h(r)}{\sqrt{\sigma^2_{i} \ast h^2(r)}},
\end{equation}
where $I_{\rm smooth}$ is the convolved image with a smoothing kernel
$h(r)$ and $\sigma^2_{\rm smooth}$ is the variance of smoothed image
that is propagated from the unsmoothed image. Given that $\chi_{\rm
smooth}$ should follow a normal distribution $N(0,1)$ for random noise,
$\chi_{\rm smooth}$ is useful to visualize the low-S/N features. Here,
we adopt a tophat kernel with a radius of 3 pixels to match the size of
apertures used for the measurements of \pol and $\theta$.
The $Q$ $\chi_{\rm smooth}$ image shows that the region with $|\chi_{\rm
smooth}| > 3$ (outlined with solid contours) is roughly aligned with the
major axis, demonstrating the significance of our polarization detections.

\begin{figure}
\epsscale{0.70}
\epsscale{1.20}
\ifpreprint\epsscale{0.90}\fi
\hspace{-0.7cm}
\plotone{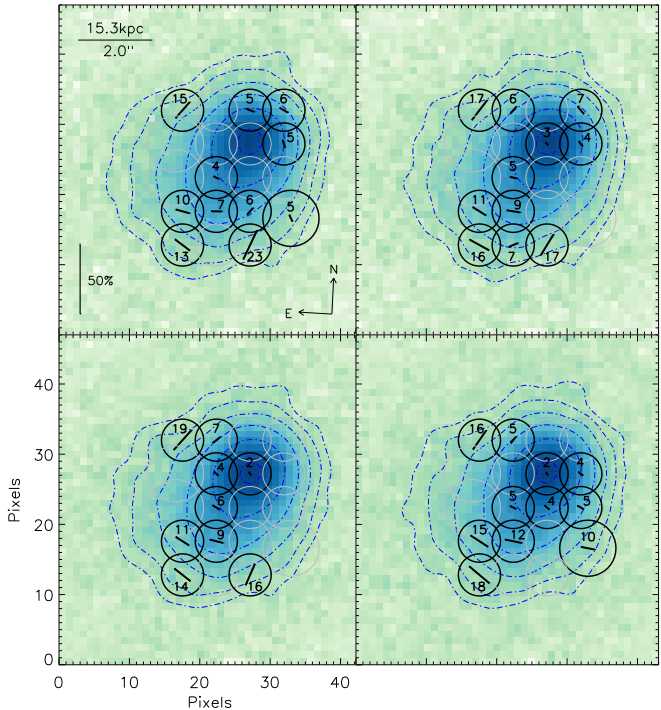}
\caption{
Four random realizations of the total polarization map after introducing
random alignment errors ($\pm$1 pixel shifts) into the process of
reducing the science images.  Their similarity to one another and to
the observed map (Fig.~\ref{fig:polmap}) suggests that our results are
robust to the positional shifts from one image to another and from one
sequence to another.
}
\label{fig:shifts} 
\end{figure}

The errors shown in Fig.~\ref{fig:polmap} are calculated purely from
photon noise statistics.  One additional source of uncertainty is
the extent to which errors in image alignment, i.e., from shifts
and rotations, affect the polarization map when we combine images.
Polarization is calculated by taking the difference between different
exposures.  When images are not aligned correctly, the polarization may
be affected.
Between sequences and within sequences, the point source in the southeast
shifts by $\sim$1 pixel and rotates relative to the blob center by only
$\sim$0.5 degree. Thus the uncertainties in misalignment are dominated
by translational errors.  To estimate how much translational errors could
affect our measurements, we introduce errors of this magnitude into our
best-aligned images and repeat the entire reduction procedure.
Figure \ref{fig:shifts} shows four random realizations of the total
polarization maps after introducing random alignment errors with $\pm$1
pixel shifts. Our results do not change significantly.

\subsection{\lya Line Polarization Map}

The UV continuum aligned with the radio lobes of radio galaxies
is sometimes polarized, with the continuum polarization fraction
\polcont typically $<10\%$, but sometimes as high as $\sim$20--30\%
\citep{Jannuzi-Elston1991, Vernet2001, Tadhunter2005}.  As a result,
the relative contributions of continuum and \lya polarization to
our total polarization map for B3 J2330+3927 are not clear.  Future
spectro-polarimetry, which could isolate the line-only polarization
signal \cite[e.g.,][]{Beck2016}, is needed.  For now, we make a
conservative argument to place lower limits on the \lya contribution,
asking whether at least some \lya polarization is required to explain
the map in Fig.~\ref{fig:polmap}.

To separate out the polarization contributed by the continuum and to
place a lower limit on the line polarization, we use the following simple
formalism, where $I_{Q, cont}$ and $I_{Q, line}$ refer to the total flux
in the $Q$ images from the continuum and \lya, respectively. This light
is polarized by $q_{cont}$ and $q_{line}$ for the continuum and \lya,
respectively.  Because the narrowband filter captures both the continuum
and \lya fluxes at the same time, in one $Q$ sequence, we measure the
total $Q$ parameter:
\begin{equation}
   \left(\frac{Q}{I}\right)_{total} = 
   \frac{ I_{Q, cont} \times q_{cont} + I_{Q, line} \times q_{line} }
        { I_{Q, cont} + I_{Q, line} }.
\end{equation}
Likewise, in a $U$-sequence, we have 
\begin{equation}
   \left(\frac{U}{I}\right)_{total} = 
   \frac{ I_{U, cont} \times u_{cont} + I_{U, line} \times u_{line} }
        { I_{U, cont} + I_{U, line} }.
\end{equation}
If we assume that the polarization angles of the continuum and \lya
are the same, using the relation
\begin{equation}
	\frac{q_{cont}}{u_{cont}} = \frac{q_{line}}{u_{line}}, 
\end{equation}
we can separate the total polarization into contributions from the
continuum and \lya:
\begin{equation}
\label{polcalclast}
	\pol = (1-f_{c})\,P_{line} + f_{c}\,P_{cont}, 
\end{equation}
where $f_c$ is the fraction of the continuum relative to the total light
captured by the narrowband filter:
\begin{equation} 
	f_{c}=\frac{I_{cont}}{I_{cont}+I_{line}}.
\end{equation}

\begin{figure}
\epsscale{1.15}
\ifpreprint\epsscale{1.00}\fi
\plotone{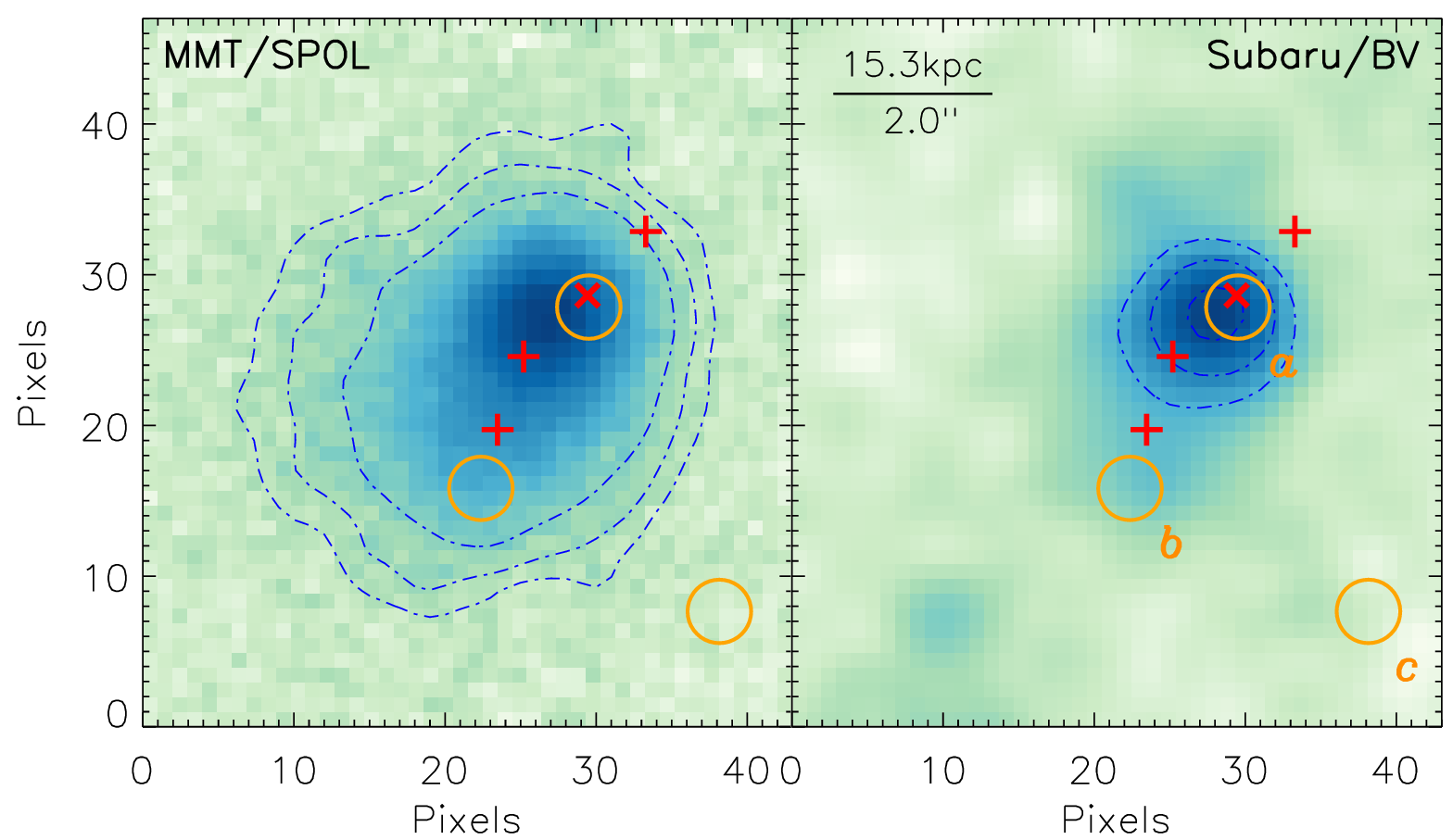}
\caption{
Map of the total (\lya plus continuum) flux within our narrowband filter
(left, as in Fig.~\ref{fig:polmap}) and the continuum flux derived
from $B$ and $V$ broadband images \citep{Matsuda2009} (right). The two
images are shown at the same stretch. The three contours represent
surface brightnesses of 0.5, 1, and 2$\times10^{-17}$ \unitcgssb.
The continuum flux contributes little to the total flux, reaching only
$\sim$10\% at blob's center and dropping off at larger radii.  As in
Fig.~\ref{fig:polmap}, the red symbols show the locations of the radio
galaxy ($\times$) and the three radio knots ($+$) in its lobes.  The orange
circles represent three $K_s$-band sources \citep{DeBreuck2003}. Source
$a$ is consistent with the AGN position, and source $c$ is a gas-rich
galaxy at the same redshift \citep{Ivison2012}.  Source $b$ is also a
companion object whose \lya is offset by +1500\,km s$^{-1}$ with respect
to source $a$ \citep{DeBreuck2003}.}
\label{fig:lyavsconti}
\end{figure}

To estimate the continuum light fraction $f_c$, we use a UV continuum
image of  B3 J2330+3927 constructed from broadband $B$ and $V$ images
\citep{Matsuda2009}, which covers a rest-frame wavelength range of 980
--1450 \AA.  Figure \ref{fig:lyavsconti} shows the SPOL (continuum + \lya)
and the Subaru (continuum) images at the same stretch.  The flux from the
\lya line dominates that from the UV continuum in our narrowband filter.
Using both the SPOL and Subaru images, we calculate $f_c$ for
the same apertures where we measured the total polarization in
Fig.~\ref{fig:polmap}.  The continuum flux, which is somewhat extended
along the radio lobe direction, is only $\sim$10\% of the total flux at
the blob's center and drops off at larger radii.

\begin{figure} 
\epsscale{0.9}
\ifpreprint\epsscale{0.50}\fi
\plotone{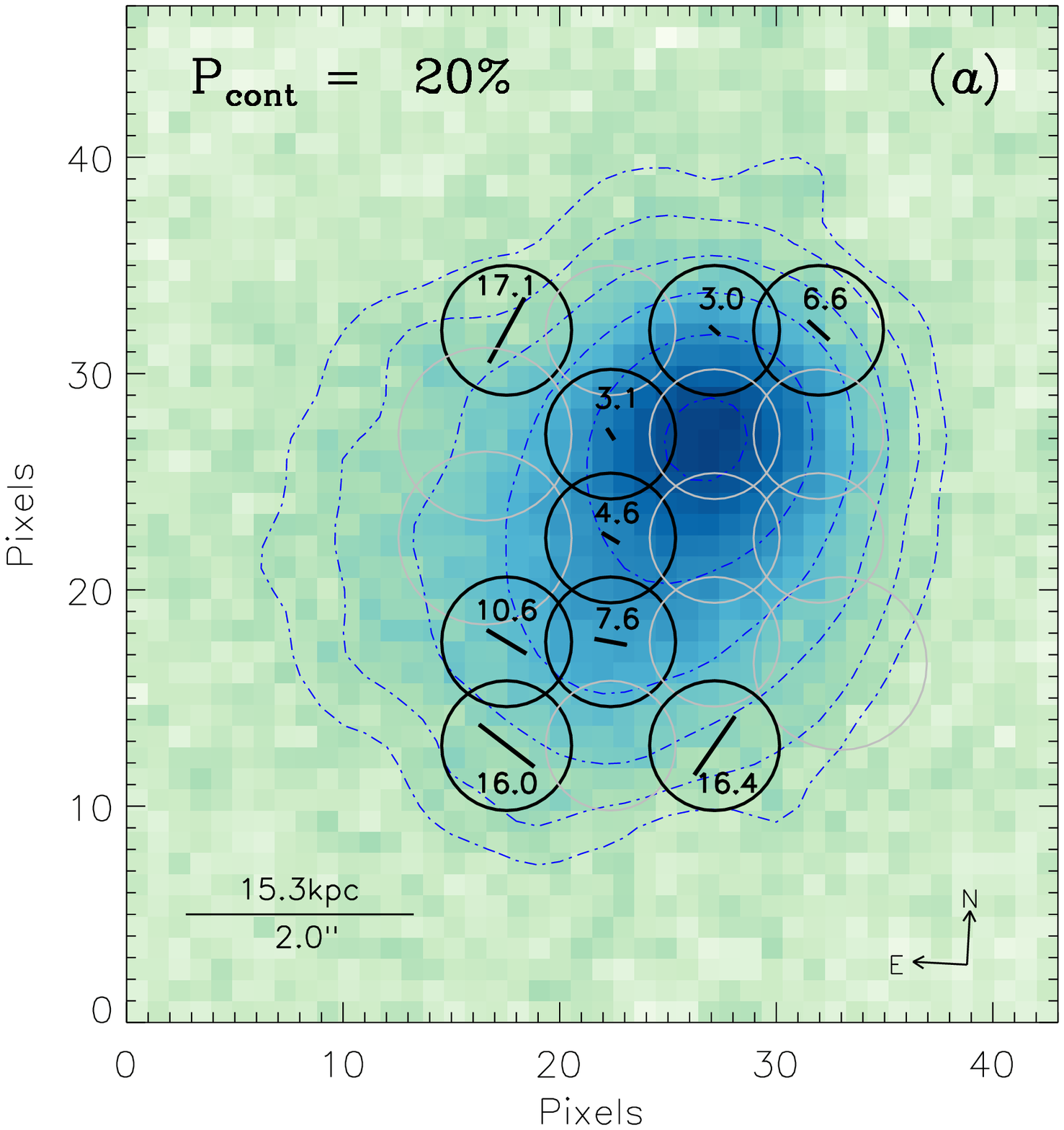} 
\plotone{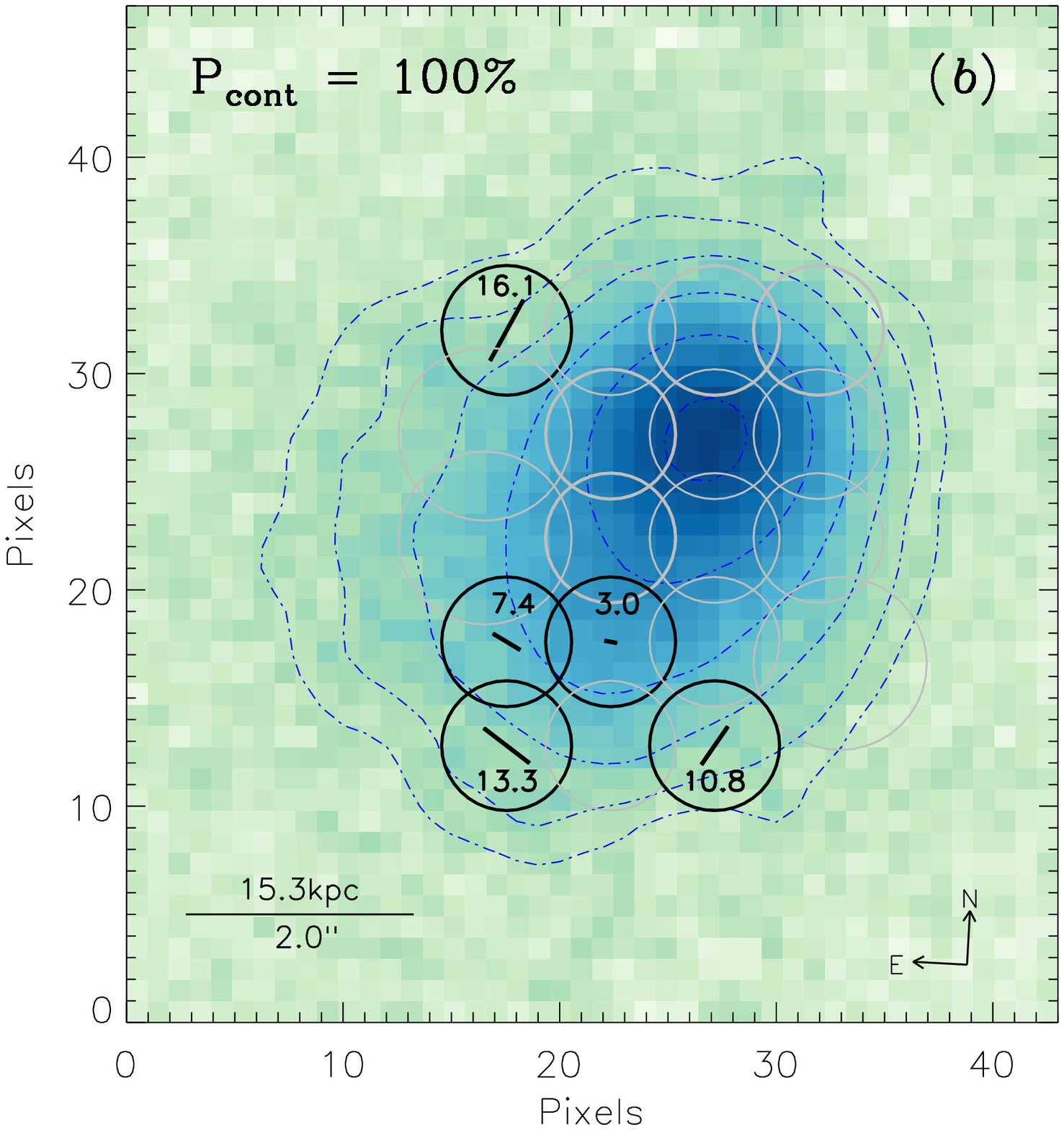} 
\caption{
Lower-limit \lya polarization maps overlaid on the total flux image.
To determine the \pollya values shown, we assume conservatively
that the continuum flux is either 20\% (panel $a$) and 100\% (panel $b$)
polarized and that its polarization angle is the same as the total
polarization.  We draw circular apertures at the same positions as in
Fig.~\ref{fig:polmap}.  For a \polcont of 20\%, consistent with some of
the highest continuum polarization fractions measured in the lobes of
radio galaxies \citep{Jannuzi-Elston1991, Vernet2001, Tadhunter2005},
\pollya contributes significantly to \pol in all nine apertures where
significant \pol is detected.  Even under the extreme and unlikely case
that \polcont is 100\%, there remain five apertures where \pollya is
still detected at $\geq 2\sigma$.
} 
\label{fig:lyapolmap} 
\end{figure}

We then consider two cases to estimate the lower limit on \pollya within
each aperture.  First, we use Eq.~(\ref{polcalclast}) to determine
\pollya under the highly conservative assumption that the UV continuum
is 20\% polarized and has a polarization direction aligned with the
total polarization.  A \polcont of 20\% is typical of the highest values
measured in radio galaxy lobes \citep{Jannuzi-Elston1991, Vernet2001,
Tadhunter2005}, Even in this case (Figure \ref{fig:lyapolmap}$a$),
\pollya contributes significantly to \pol in all nine apertures where
significant \pol is detected.  The \pollya values here range from 3
to 17\% at $\sim$10--25\,kpc, with no significant \lya polarization
detected near the blob center.  Like the total polarization, the \lya
polarization detections occur more often along the blob's major axis.
If we assume instead that \polcont is 100\% (panel $b$), an assumption
so extreme that it requires negative (unphysical) \pollya values for
many apertures given \pol, there remain five apertures in the southeast
where \pollya is still detected at $\geq 2\sigma$.

\subsection{Physical Interpretation}

From the first \lya imaging polarimetry of a radio-loud \lya nebula,
B3 J2330+3927, we find that the total polarization fraction \pol
increases from $<$2\% at the blob center to 17\% at $\sim$15--25 kpc.
Significant polarization is detected preferentially along the blob's
major axis at angles perpendicular to that axis.  In this section, we
briefly discuss the implication of our measurements. Future papers will
focus on detailed comparisons with numerical models.

Imaging polarimetry is a useful tool to differentiate between a central
powering geometry and an extended power source.  In the former case,
\lya photons are produced by a central point-source or sources (i.e.,
embedded star-forming galaxies or AGN) and transported to large radii.
When the central source illuminates the surrounding neutral gas,
the \lya photons do not experience much resonant or core scattering
and escape the system via Rayleigh or wing scattering.  The resultant
\lya radiation is highly polarized at large radii and the polarization
angle is aligned tangentially to the overall geometry of the system
\citep{Rybicki&Loeb1999, Dijkstra-Loeb2008}.
In contrast, in the latter case of extended emissivity, \lya photons
are produced {\it in situ} in the extended gas through hydrogen
recombination following ionization by photo-ionizing sources (e.g.,
AGNs) or superwind-driven shock-heating.  Because the \lya photons have
no preferential orientation with respect to the neutral medium and the
observers, little or no polarization is expected.

In B3 J2330+3927, the observed high \lya polarization fraction ($\sim$20\%
at the largest radii) and extended continuum emission suggest that \lya
photons are produced in the center, instead of arising throughout the
nebula itself.

Likewise, the observed increase in polarization with radius is consistent
with theoretical predictions from \citet{Dijkstra-Loeb2008} assuming a
simple geometry and central source.  In their expanding shell model,
the \lya polarization gradient arises when photons at larger radii
scatter by larger angles (closer to 90\degr) toward the observer.
In an alternative model of an optically thick, spherically symmetric,
collapsing gas cloud, the \lya radiation field becomes more anisotropic
at larger radii. In other words, photons tend to propagate more radially outward prior
to their last scattering events, requiring a larger scattering angle to
reach the observer, and thus are more polarized.

While some of the polarization properties of B3 J2330+3927 (fractions,
angles, and radial gradient) are qualitatively similar to those of
SSA22-LAB1 \citep{Hayes2011}, one difference is that the significant
polarization favors the major axis (and radio-jet direction).
We speculate that the lack of polarization detected along the minor axis
could be due to strong obscuration from an AGN torus perpendicular to
the radio-jet.  Another possibility is that ionization states and optical
depths vary from one axis to the other due to photo-ionization along the
jet or its interaction with the IGM.  In this case, \lya photons can
escape the system with fewer scatterings in the major axis direction.
It is not known whether this polarization pattern is common for other
giant \lya nebulae around high-$z$ radio galaxies.  To investigate
these issues further, we need deeper and higher spatial resolution
observations of this system and a systematic survey of polarization for
a larger sample.

\section{Conclusions}
\label{sec:conclusion}

We present the first narrow-band, imaging polarimetry of a \lya nebula, or
``blob," with an embedded, radio-loud AGN.   The blob, B3 J2330+3927, lies
at $z=3.09$, extends over $\sim$150\,kpc, and has a \lya luminosity of
2.5 $\times$ $10^{44}$ \ergs\ \citep{DeBreuck2003, Matsuda2009}.  The AGN
lies near the \lya emission peak and its radio lobes align roughly with
the major axis of the blob's extended \lya emission.  Our findings are:

\begin{enumerate}[leftmargin=0.5cm]

\item  We map the total (\lya plus continuum) polarization in a grid of
circular apertures of radius $0.6^{\prime\prime}$ (4.4\,kpc), detecting
significant ($\geq 2\sigma$) polarization in nine apertures and achieving
strong upper-limits (as low as 2\% in the total polarization fraction
\pol) elsewhere.

\item  The gradient in the total polarization map increases from \pol
$<2$\% at $\sim$5\,kpc from the blob center to 17\% at $\sim$15--25 kpc.
The detections lie mostly along the blob's major axis and the polarization
angles are generally perpendicular to it.

\item  Comparing the total flux to that of the continuum, and assuming
conservatively that the continuum is 20--100\% polarized and aligned with
the total polarization, we place lower limits on the \lya polarization
fraction \pollya.  Under these assumptions, \pollya is 3--17\% at
$\sim$10--25\,kpc.  No significant \lya polarization is detected at
$\sim$5\,kpc of the blob center.  Like the total polarization, the \lya
polarization detections tend to lie along the blob's major axis.

\end{enumerate}

Our polarization measurements for B3 J2330+3927 complement past
polarization work, which focused on radio-{\it quiet} blobs and on
radio galaxies within \lya clouds.  For example, the polarization of
SSA22--LAB1 is not measurable at its center, but rises to $\sim$10\%
at $\sim$30\,kpc and to $\sim$20\% at $\sim$45\,kpc, forming an almost
complete polarized ring \citep{Hayes2011}.  While the polarization that
we detect in B3 J2330+3927 is also tangentially-oriented and outside
the blob center (and AGN), it is generally significant only along the
blob's major axis (and radio lobe direction).

Unlike previous studies, we have constrained and mapped the \lya
contribution to the total polarization. The one spectro-polarization
measurement isolating the \lya line in a radio-loud \lya blob also reveals
its polarization fraction to be high (16\%) and perpendicular to the radio
lobe axis in a region 10--40\,kpc from the nucleus, at least on one side
of the nebula \citep{Humphrey2013}.  Such a high \pollya has not been
observed in radio galaxies \cite[e.g.,][]{Dey1997,Cimatti1998}, which
might imply a physical difference or arise from \pollya being measured on
smaller physical scales. Spatially-resolved measurements of \pollya for
a larger sample of radio galaxies are required to discriminate between
these scenarios.

A direct comparison of our narrow-band, imaging polarimetry in B3
J2330+3927 with our on-going survey of \lya blobs without known AGN
and with radio-quiet AGN will improve greatly our understanding of the
mysterious source of their extended \lya emission.

\acknowledgements

We thank the referee, Matthew Hayes, for his thorough reading of the
manuscript and helpful comments.
We thank the staff at the MMT Observatory for their efforts in support of
this program.  
We thank Daryl Willmarth and the NOAO for making the narrowband filter
available for our observations.
C.Y. and A.I.Z. acknowledge support from the NSF Astronomy
and Astrophysics Research Program through grant AST-0908280 and from
the NASA Astrophysics Data Analysis Program through grant NNX10AD47G.
Y.Y. and E.K.'s research was supported by Basic Science Research Program
through the National Research Foundation of Korea (NRF) funded by the
Ministry of Science, ICT \& Future Planning (NRF-2016R1C1B2007782).
Y.Y. acknowledges support from the BMBF/DLR grant Nr.\ 50 OR 1306.
M.K.M.P. was supported by a Dark Cosmology Centre Fellowship. The Dark
Cosmology Centre was funded by The Danish National Research Foundation.
M.G.L. and E.K. are supported by the National Research Foundation of Korea
(NRF) grant funded by the Korea Government (MSIP) (No.~2012R1A4A1028713).

\medskip
\ifpreprint
	\facility{MMT (SPOL)}
\else
	Facilities: 
	\facility{MMT (SPOL)}
\fi

\end{document}